\input harvmac
\input epsf

\overfullrule=0pt
\abovedisplayskip=12pt plus 3pt minus 3pt
\belowdisplayskip=12pt plus 3pt minus 3pt
%

\def\bar{\overline}

\def\cN{{\cal N}}

\def\bigone{\hbox{1\kern -.23em {\rm l}}}
\def\ZZ{\hbox{\zfont Z\kern-.4emZ}}

\font\zfont = cmss10 

\def\bigone{\hbox{1\kern -.23em {\rm l}}}
\def\ZZ{\hbox{\zfont Z\kern-.4emZ}}


\lref\doa{ M. R. Douglas, G. Moore, {\it D-branes,
Quivers, and ALE
Instantons}, hep-th/9603167.}
\lref\dob{M. R. Douglas, B. R. Greene, D. R. Morrison,
{\it Orbifold
Resolution by D-Branes},  hep-th/9704151,  Nucl.Phys.
{\bf B506}
(1997) 84.}
\lref\mala{J. Maldacena, {\it  The Large N Limit of
Superconformal
Field Theories and Supergravity}, hep-th/9711200,
Adv.Theor.Math.Phys. {\bf 2}
(1998) 231-252.}
\lref\malb{O. Aharony, S.S. Gubser, J. Maldacena, H.
Ooguri, Y. Oz,
{\it Large N Field Theories, String Theory and
Gravity}, 
hep-th/9905111,
Phys.Rept. {\bf 323} (2000) 183.}
\lref\kac{S. Kachru, E. Silverstein, {\it 4d
Conformal Field Theories
and Strings on Orbifolds}, hep-th/9802183, 
Phys.Rev.Lett. {\bf 80} 
(1998)
4855.}
\lref\law{A. Lawrence, N. Nekrasov, C. Vafa, {\it On
Conformal
Theories in Four Dimensions}, hep-th/9803015,
Nucl.Phys. {\bf B533} 
(1998)
199.}
\lref\senreview{A. Sen, {\it Non-BPS States and
Branes in String
Theory}, hep-th/9904207.}
\lref\senworld{A. Sen, {\it Supersymmetric
World-Volume Action For
Non-BPS D-Branes}, hep-th/9909062, JHEP {\bf 10}
(1999) 008.} 
\lref\sentach{A. Sen, {\it Tachyon Condensation on
the Brane
Anti-Brane System}, hep-th/9805170; JHEP {\bf 08}
(1998) 012.}
\lref\senuniv{A. Sen, {\it Universality of the
Tachyon Potential}, hep-th/9911116, 
JHEP {\bf 12} (1999) 027.}
\lref\senbound{A. Sen, {\it Stable Non-BPS Bound
States of 
BPS D-branes}, hep-th/9805019, JHEP {\bf 08} (1998)
010.}
\lref\sencycle{A. Sen, {\it BPS D-Branes on
Nonsupersymmetric
Cycles}, hep-th/9812031, JHEP {\bf 12} (1998) 021.}
\lref\joysen{J. Majumder and A. Sen, {\it Blowing
Up D-Branes 
on Nonsupersymmetric Cycles}, hep-th/9906109, JHEP
{\bf 09} (1999)
004.} 
\lref\gabsen{M. Gaberdiel and A. Sen, {\it
``Nonsupersymmetric
D-Brane Configurations with Bose-Fermi Degenerate Open
String
Spectrum''}, hep-th/9908060; JHEP {\bf 11} (1999)
008.}
\lref\naraintach{E. Gava, K.S. Narain and M.H.
Sarmadi, {\it ``On the 
Bound States of p-Branes and (p+2)-Branes''},
hep-th/9704006;
Nucl. Phys. {\bf B504} (1997) 214.}
\lref\piljin{P. Yi, {\it ``Membranes from Five-Branes
and Fundamental
Strings from Dp-Branes''}, hep-th/9901159; 
Nucl. Phys. {\bf B550} (1999) 214.}
\lref\polkthree{J. Polchinski, {\it ``Tensors from K3
Orientifolds''},
hep-th/9606165; Phys. Rev. {\bf D55} (1997) 6423.}
\lref\dmfrac{K. Dasgupta and S. Mukhi, {\it Brane
Constructions, 
Fractional Branes and Anti-de Sitter Domain Walls},
hep-th/9904131,
JHEP {\bf 07} (1999) 008.}
\lref\dmconif{K. Dasgupta and S. Mukhi, {\it Brane
Constructions,
Conifolds and M-Theory}, hep-th/9811139, Nucl. Phys.
{\bf B551} 
(1999) 204.}
\lref\urangaconif{A. Uranga, {\it Brane
Configurations for Branes at
Conifolds}, hep-th/9811004, JHEP {\bf 01} (1999)
022.}
\lref\dougmoore{M. Douglas and G. Moore, {\it D
Branes, Quivers and 
ALE
Instantons}, hep-th/9603167.}
\lref\fracbranes{M. Douglas, {\it Enhanced Gauge
Symmetry in M(atrix)
Theory}, hep-th/9612126, JHEP {\bf 07} (1997)
004 \semi 
D.-E. Diaconescu, M. Douglas and J. Gomis, {\it
Fractional Branes and 
Wrapped Branes}, hep-th/9712230, JHEP {\bf 02}
(1998) 013.} 
\lref\kafra{A. Karch, D. L\"ust and D. Smith, {\it
Equivalence of 
Geometric Engineering and Hanany-Witten via Fractional
Branes},
hep-th/9803232, Nucl. Phys. {\bf B533} (1998) 348.}
\lref\kenwilk{C. Kennedy and A. Wilkins, {\it
Ramond-Ramond
Couplings on Brane-Antibrane Systems},
hep-th/9905195;
Phys. Lett. {\bf B464} (1999) 206.}
\lref\gabstef{M. Gaberdiel and B. Stefanski, {\it
Dirichlet  Branes 
on
Orbifolds}, hep-th/9910109.}
\lref\klebwit{I. Klebanov and E. Witten, {\it
Superconformal Field 
Theory on Threebranes at a Calabi- Yau Singularity},
Nucl.Phys.
{\bf B536} (1998) 199, hep-th/9807080.}
\lref\malda{J. Maldacena, {\it The Large N Limit of
Superconformal Field Theories and Supergravity},
Adv. Theor. Math. Phys. {\bf 2} (1998) 231;
hep-th/9711200.}
\lref\gubkleb{S. S. Gubser and I. R. Klebanov, {\it
Baryons and 
Domain
Walls in an $\cN =1$ Superconformal Gauge Theory},
Phys. Rev. {\bf D58} (1998) 125025, hep-th/9808075.}
\lref\klenek { I. R. Klebanov, N. A. Nekrasov, {\it
Gravity Duals of 
Fractional Branes and Logarithmic RG Flow},
hep-th/9911096, Nucl.Phys. {\bf B574} (2000) 263.}
\lref\kletse { I.R. Klebanov, A.A. Tseytlin, {\it
Gravity Duals of 
Supersymmetric $SU(N) \times SU(N+M)$ Gauge Theories},
hep-th/0002159, Nucl.Phys. {\bf B578} (2000) 123.}  
\lref\gns{S. Gubser, N. Nekrasov and S. Shatashvili,
{\it
Generalized Conifolds and Four Dimensional $\cN=1$
Superconformal
Theories}, hep-th/9811230, JHEP {\bf 05} (1999) 003.}
\lref\ksil{S. Kachru and E. Silverstein, {\it 4D
Conformal Field
Theories and Strings on Orbifolds}, Phys. Rev. Lett.
{\bf 80}
(1998) 4855; hep-th/9802183.} 
\lref\lnv{A. Lawrence, N. Nekrasov and C. Vafa, {\it
On Conformal
Field Theories in Four Dimensions}, Nucl. Phys. {\bf
B533} (1998)
199; hep-th/9803015.}
\lref\seiwitnc{N. Seiberg and E. Witten, {\it String
Theory and 
Noncommutative Geometry}, hep-th/9908142, JHEP {\bf 09} (1999) 032.}
\lref\ot{ K. Oh and R. Tatar, {\it Renormalization
Group Flows on D3 
branes at an Orbifolded Conifold}, hep-th/0003183,
JHEP {\bf 05} (2000) 030.}
\lref\karchmina{M. Aganagic, A. Karch, D. Lust and A.
Miemec, {\it 
Mirror Symmetries for Brane Configurations and Branes
at 
Singularities},hep-th/9903093,
Nucl. Phys. {\bf B569} (2000) 277.}
\lref\ota{ K. Oh, R. Tatar, {\it Branes at Orbifolded
Conifold 
Singularities
 and Supersymmetric Gauge Field
Theories}, hep-th/9906012, JHEP {\bf 10} 
(1999) 031.}   
\lref\senworld{A. Sen, {\it Supersymmetric
World-Volume Action For
Non-BPS D-Branes}, hep-th/9909062, JHEP {\bf 10}
(1999) 008.} 
\lref\panda{E. A. Bergshoeff, M. de Roo, T. C. de Wit,
E. Eyras and S. 
Panda,
 {\it ``T-Duality and Actions for Non-BPS D-Branes''},
hep-th/0003221.}
\lref\garou{M. R. Garousi, {\it ``Tachyon Coupling on
Non-BPS D-Branes 
and Dirac-Born-Infeld Action''}, hep-th/0003122.} 
\lref\minstro{R. Gopakumar, S. Minwalla and A.
Strominger, 
{\it Noncommutative
Solitons}, hep-th/0003160, JHEP {\bf 05} (2000) 020. }
\lref\bcr{M. Bill\`o, B. Craps and F. Roose, {\it
Ramond-Ramond 
Coupling
of Non-BPS D-Branes}, hep-th/9905157; JHEP {\bf 06}
(1999) 033. }
\lref\dmr{K. Dasgupta, S. Mukhi and G. Rajesh, {\it
Noncommutative
Tachyons}, hep-th/0005006, JHEP {\bf 06} (2000) 022.}
\lref\hkl{J. A. Harvey, P. Kraus, F. Larsen, E. J.
Martinec, 
{\it Strings and Branes as Noncommutative
Solitons}, 
hep-th/0005031, JHEP {\bf 07} (2000) 042.}
\lref\witn{E. Witten, {\it Noncommutative Tachyons
And String Field
Theory},  hep-th/0006071.}
\lref\witm{E. Witten, {\it Solutions Of
Four-Dimensional Field
Theories Via M Theory},  hep-th/9703166, 
Nucl.Phys. {\bf B500} (1997) 3.}
\lref\senas{A. Sen,{\it Descent Relations Among
Bosonic D-branes}, 
hep-th/9902105, 
Int. J. Mod. Phys {\bf A 14} (1999) 4061.}
\lref\sena {A. Sen and B. Zwiebach, {\it Tachyon
Condensation in String 
Field Theory},
hep-th/9912249, JHEP {\bf 03} (2000) 002.}
\lref\tay{ W. Taylor, {\it D-brane Effective Field
Theory from String 
Field Theory}, hep-th/0001201.}
\lref\tayl{N. Moeller and W. Taylor, {\it Level
Truncation and the 
Tachyon in Open Bosonic SFT},
hep-th/0002237, Nucl.Phys. {\bf B583} (2000) 105.}
\lref\ghse{ D. Ghoshal, A.  Sen, {\it Tachyon
Condensation and Brane 
Descent Relations in p-adic String Theory},
hep-th/0003278, Nucl.Phys. {\bf B584} (2000) 300.}
\lref\hk { J. A. Harvey and P. Kraus, {\it D-Branes
and Lumps in 
Bosonic Open String Field Theory},
hep-th/0002117, JHEP {\bf 04} (2000) 012.}
\lref\djmt{R. de Mello Koch, A. Jevicki, M. Mihailescu
and R. Tatar 
{\it Lumps and P-branes in Open String 
Field Theory}, hep-th/0003031, Phys. Lett. {\bf B482} (2000) 249.}
\lref\berk{N. Berkovits, {\it The Tachyon Potential in
Open 
Neveu-Schwarz String Field Theory}, hep-th/0001084,
JHEP {\bf 04} (2000) 022.}
\lref\berko{N. Berkovits, A. Sen and B. Zwiebach, {\it
Tachyon 
Condensation in Supestring Field Theory},
hep-th/0002211.}
\lref\berkov{ P. J. De Smet, J. Raeymaekers, {\it
Level Four 
Approximation to the Tachyon Potential in 
Superstring Field Theory},  hep-th/0003220, JHEP {\bf 05} (2000) 051.}
\lref\berkovi{ A. Iqbal, A. Naqvi, {\it Tachyon
Condensation on a 
non-BPS D-brane}, hep-th/0004015.}
\lref\berkovit{N. Moeller, A. Sen, B. Zwiebach, {\it
D-branes as 
Tachyon Lumps in String Field Theory}, hep-th/0005036, JHEP {\bf 08} (2000) 039 }
\lref\berkr{R. de Mello Koch, J. P. Rodrigues, {\it
Lumps in level 
truncated open string field theory},
hep-th/0008053.}
\lref\berkm{N. Moeller, {\it Codimension two lump
solutions in
string field theory and tachyonic theories},
hep-th/0008101.}
\lref\zwiu{L. Rastelli and B. Zwiebach, {\it Tachyon potentials, 
star products and universality}, hep-th/0006240.}
\lref\zwid{A. Sen and B. Zwiebach, {\it Large Marginal Deformations in 
String Field Theory}, hep-th/0007153.}
\lref\zwit{A. Iqbal and A. Naqvi, {\it On Marginal Deformations in 
Superstring Field Theory}, hep-th/0008127}
\lref\zwip{B. Zwiebach, {\it A Solvable Toy Model for 
Tachyon Condensation in String Field Theory}, hep-th/0008227.}
\lref\hm{ J. A. Harvey, G. Moore, {\it Noncommutative Tachyons and K-Theory},
hep-th/0009030.}
\lref\dhot {K. Dasgupta, S. Hyun, K. Oh , R. Tatar,
{\it Conifolds with
Discrete Torsion and
Noncommutativity}, hep-th/0008091, to appear in JHEP.} 
\lref\hakl{ J. A. Harvey, P. Kraus, F. Larsen, 
{\it Tensionless Branes and Discrete Gauge Symmetry},
hep-th/0008064.}
\lref\mar{G. Mandal, S.-J. Rey, {\it A Note on D-Branes of Odd Codimensions 
from Noncommutative Tachyons}, hep-th/0008214.}
\lref\gmst{R. Gopakumar, S. Minwalla, A. Strominger, {\it Symmetry 
Restoration and Tachyon Condensation in Open String Theory},  hep-th/0007226.}
\lref\soc{C. Sochichiu, {\it Noncommutative Tachyonic Solitons. Interaction 
with Gauge Field}, hep-th/0007217, JHEP {\bf 08} (2000) 026.}
\lref\seib{N. Seiberg,{\it A Note on Background Independence in 
Noncommutative Gauge Theories, Matrix Model and Tachyon Condensation}, 
hep-th/0008013, JHEP {\bf 09} (2000) 003.}
\lref\yi{P. Yi, {\it Membranes from Fivebranes and
Fundamental Strings 
from Dp Branes}, hep-th/9901159, Nucl. Phys. {\bf B550}
(1999) 214.}
\lref\sentach{ A. Sen, {\it Supersymmetric Worldvolume
Action for Nonbps D-Brane}, hep-th/9909062, JHEP {\bf 9910} (1999).}
\lref\hori{O. Bergman, K. Hori, P.Yi, {\it Confinement
on the Brane},
hep-th/0002223, Nucl. Phys. {\bf B580} (2000) 289.}
\lref\pru{ J. Park, R. Rabadan, A. M. Uranga, {\it
Orientifolding
the conifold},   hep-th/9907086, Nucl.Phys. {\bf
B570} (2000) 38}
\lref\ms{S. Mukhi and N. V. Suryanarayama, {\it Chern-Simons Terms 
on Noncommutative Branes}, hep-th/0009101.} 
\lref\kw{C. Kennedy and A. Wilkins, {\it Ramond-Ramond
Couplings on 
Brane-Antibrane Systems}, hep-th/9905195, Phys.Lett. {\bf B464} (1999) 
206}
\lref\msa{S. Mukhi, N. V. Suryanarayana, D. Tong, {\it
 Brane-Antibrane 
Constructions}, hep-th/0001066,  JHEP {\bf 03}
(2000) 015}
\lref\msb{ S. Mukhi, N. V. Suryanarayana,{\it A Stable
Non-BPS Configuration From Intersecting Branes and
Antibranes}, hep-th/0003219, JHEP {\bf 
06} (2000) 001 }
\lref\sens{A. Sen, {\it Some Issues in Non-commutative
Tachyon 
Condensation}, hep-th/0009038.}
\lref\sense{A. Sen, {\it Uniqueness of Tachyonic
Solitons}, 
hep-th/0009090.} 
\lref\otw{Y. Oz, T. Pantev, D. Waldram, {\it
Brane-Antibrane Systems on
Calabi-Yau Spaces}, hep-th/0009112}
\lref\kleb{I. R. Klebanov, {\it TASI Lectures: Introduction to the 
AdS/CFT Correspondence}, hep-th/0009139.} 
\lref\dom{M. R. Douglas, G. Moore, {\it D-branes,
Quivers, and ALE Instantons}, hep-th/9603167} 
\Title{\vtop{\hbox{hep-th/0009213}
\hbox{BROWN-HET-1240}
\hbox{HU-EP-00/33}}}
{\vbox{\centerline{A Note on Non-Commutative Field Theory}
\bigskip
\centerline{and Stability of Brane-Antibrane Systems}}}
\centerline{Radu Tatar}
\vskip 5pt
\centerline{\it Humboldt-Universit\"at zu Berlin,
Institut f\"ur  
Physik,}
\centerline{\it Invalidenstrasse 110, 10115 Berlin,
Germany \foot{Permanent Address after September 1, 2000}}
\centerline{and}
\centerline{\it Department of Physics, Brown
University, Providence, RI
02912, USA.}

\ \smallskip
\centerline{ABSTRACT}

It has been conjectured that a pair of
 $D5 - \bar{D5}$ branes 
wrapped on  some non-trivial two cycle of a 
Calabi-Yau manifold  becomes a stable BPS $D3$ brane 
in the presence of a very large $B$ field
and magnetic fluxes on their worldvolumes. 
We discuss this by considering the non-commutative
field theory on the worldvolume of the pair of branes whose field 
multiplication is made with respect to two different $*$ products 
due to the presence of different  
$F$ fields on the two branes. The tachyonic field becomes massless for a 
specific choice of the magnetic fluxes and it allows a trivial solution.
Our discussion generalizes recent results concerning stability of 
brane-antibrane systems on Calabi-Yau spaces to the case of 
non-commutative branes.
\Date{September 2000}
\vfill\eject
\ftno=0

\newsec{Introduction}
In the last few years an important amount of work has
been done for
studying D branes at orbifold and conifold sigularities 
\refs{\doa,\dob}
and their conformal limit within the AdS/CFT
correspondence
\refs{\mala,\malb} as discussed in 
\refs{\kac,\law,\klebwit, \urangaconif, \dmconif, \karchmina,
\ota, \dhot, \kleb}.

If we consider four dimensional field theories, they can appear either on 
 $D3$ branes which are orthogonal to the singular space and carry integer 
charge or on (anti) $D5$ branes wrapped 
on different vanishing 2-cycles arising in 
the resolution of the singularity in which case the $D3$ branes 
are fractional branes and carry charges measured in terms of the $B_{NSNS}$ 
fluxes on the resolution 2-cycles \refs{\fracbranes, \kafra, \dmfrac}.

In this paper we will concentrate on the
phenomena appearing on a
pair of $D5 - \bar{D5}$ branes 
wrapped on $S^2$ cycles in the presence of a large
B field and different fluxes on their worldvolumes.
The two fractional branes have tension and charge
proportional to $B$ and $(1-B)$, the constant being
understood to arise from
a world-volume magnetic flux that must be turned on in
the antibrane and keeping the flux on the $D5$ brane equal to zero 
\refs\dmfrac.

For general brane-antibrane pairs, there is a tachyon and the pairs
are generically unstable \refs\senreview.
The potential for the tachyon is an universal
function with overall multiplicative factors coming from the brane
tension \refs\senuniv. For any nonzero value of the $B$-flux through
the $S^2$ cycle, one may therefore naively expect to find a
tachyon with its associated potential. However, we
know from \refs\msa\ that for a
special choice of background $B$ and $F$ fields the tachyon becomes a
massless scalar field and this implies that one should have a
stable system. In the paper \refs\dmfrac\ it has been conjectured that this is
 an integer  BPS $D3$-brane. The problem is to explain the 
fate of the massless scalar which is not seen in the field theory on the 
integer lower $D3$-brane.
We are going to argue in the present paper that there
is a trivial solution for the massless scalar field which 
actually therefore does not appear in the field theory
on the brane-antibrane pair. The trivial solution will also determine the
fact that the $U(1) \times U(1)$ gauge group survives. 

The content of the present paper is as follows.
In section 2 we will describe the non-commutative field theory on a
 $D5 - \bar{D5}$ pair wrapped on a vanishing $S^2$ cycle, 
in the presence of large $B$ and different $F$  fields on the worldvolumes of
the pair components. We use the 
assumption of holomophicity for the fields \refs\otw\ to describe a trivial
solution for the tachyon. This solution for the tachyon is different from the 
one of \refs\hkl\ where a B field is present but 
the magnetic fluzes on the branes are equal and from
the one of reference 
\refs\otw\ where the magnetic fluxes are different but there is no
B field and the field theory is commutative. We discuss a smooth transition 
between the two previous known solution, transition which neccesarily passes
through our solution.

The result is that the pair of  $D5 - \bar{D5}$ cancel each other but what
remains is an integer $D3$ brane instead of a tachyon condensation. One 
question that arises here is how could this happen in the view of the 
universality of the tachyon potential which tells us that the same phenomenon
happens in any background. The answer is that the universality argument holds
only for tachyons with zero momentum. In the case of different fluxes on the 
pair of  $D5 - \bar{D5}$ branes, the tachyon are charged and they do not have
zero momentum and the universality argument does not work. It would be very
interesting to have a general description of this phenomenon. 

In section 3 we discuss our results in connection with the results of
\refs\dmfrac\ concerning the field theory on  $D5 - \bar{D5}$ pair 
wrapped on the vanishing $S^2$ cycle at the apex of a conifold singularity,
when we have a unit flux on the $\bar{D5}$.

\newsec{Fluxes and Stability of $D5- \bar{D5}$ wrapped
on an $S^2$ 
cycle}

Recently, important evidence has been accumulated
showing that 
string field theory provides a direct approach to
study string theory
tachyons. The tachyons of  unstable systems such 
as non-BPS D-branes or pair of
brane-antibrane acquire an expectation value at a
minimum of their
potential where the total negative potential energies
exactly cancel the
tensions of the unstable systems 
\refs{\senbound,\senworld,\senas,\sencycle}.
These conjectures were checked by using approximation
schemes in open
string field theory and the acumulated evidence is
impressive 
\refs{\senuniv,\sena,\tay,\tayl, \hk, \djmt, \berk,
\berko, \berkov,
\berkovi, \berkovit, \berkr, \berkm,\zwiu,\zwid,\zwit,\zwip}.
In the framework of
\refs{\minstro,\dmr,\hkl,\witn,\soc,\gmst,\seib,\mar,\sens,\hm,\sense}, 
the description of the tachyon
condensation is much simpler when a large background $B$ field is
turned on. In our case a supplementary feature is that, besides
the B field, we also turn on different fluxes on different branes.
This will determine a different solution in our case for specific
values of fluxes on pairs of branes and antibranes.

Consider the case when we have a flux $F_1$ on the worldvolume of the
 $D5$ brane and a flux $F_2$ on the worldvolume of the $\bar{D5}$ brane.
Due to the presence of different gauge
fields on the two branes together with the $B$ field, the field theory on the
 worldvolumes of the pair of branes becomes non-commutative and 
we have two different *-products which are parametrised by \refs\seiwitnc:
\eqn\tt{\theta_I^{ij}={\Big(}{1\over F_I-B}
{\Big)}\epsilon^{ij} \equiv \theta_I,~~I=1,2}
The field content of the theory is made of a $U(1)$ gauge boson $A_{1 \mu}$
on the $D5$ brane, a $U(1)$ gauge boson $A_{2 \mu}$ on the $\bar{D5}$ brane
and a tachyon $T$ which is charged under both $U(1)$ groups. 

The effective theory is a $U(1)\times U(1)$ non-commutative gauge
theory and the fields are multiplied by a noncommutative $*_{I}$ product
with noncommutativity parameter $\theta^I,~(I=1,2)$
given as above. The gauge field $A_{1 \mu}$ is multiplied 
by  $*_{I}$ product, the gauge field $A_{2 \mu}$ is multiplied 
by  $*_{I}$ product respectively. The  complex tachyon $T=T_{12}$ 
together with its complex
conjugate ${\bar T}=T_{21}$ being charged with respect to both
gauge groups  
are also multiplied by using one of the $*_{I}$ products: 
namely, $T_{12}*T_{21}$ is defined by using $*_2$ while
$T_{21}*T_{12}$
is defined by using $*_1$. For a $D5-\bar{D5}$ system,
the tachyon is a 
 $2 \times 2$ matrix, the $*$ product appears as $2 \times 2$ matrix too and
the  $*$ products are applied to matrices. 
Because of the associativity of matrix products, it 
results that the $*$ product is associative and this is one of the required 
properties for any $*$ product.

A general discussion of field theories with different $*_{I}$ products 
would be very
interesting and could reveal many important results but 
at this point let us concentrate on a  $D5-\bar{D5}$
system wrapped on a vanishing $S^2$ cycle. The fact
that the cycle is a vanishing one determines a large value for the $B$ and 
 $F$ field in order to obtain a finite flux, and this 
implies a large non-commutativity condition 
Because there is no tachyon in the spectrum of the
strings with both
ends on the $D5$ brane or both ends on the $\bar{D5}$
brane but on the spectrum of the open strings which connect the pair,  
the tachyon is given by a $2 \times 2$ matrix
\eqn\tamax{
T = \pmatrix{0&T \cr T^*&0}}
i.e. $T_{12} = T , T_{21} = T^*$.

With $B$ field and $F$ fields on the world
volumes of the pair of branes, the low-energy effective action is
\eqn\action{ S = \int_{S^2} \left[{1 \over 4} Tr F_1^2 + 
{1 \over 4} Tr_2 F_2^2  +  DT * DT^* + V(T, T^{*})\right] \ .}
In the formula for the
effective action the gauge
field strenghts have the canonical form and the
covariant derivatives 
are
given by:
\eqn\cova{D_{\mu} T = \partial_{\mu} T + i (A_{1 \mu}
*_{1} T 
        - i T *_{2} A_{2 \mu}) \ .}
and
\eqn\covb{D_{\mu} T^* = \partial_{\mu} T^* + i ( T^*
*_{1} A_{1 \mu}  
        - i A_{2 \mu} *_{2} T^{*}) \ .}
The $V(T, T^{*})$ is the form of the tachyonic
potential and this can 
be extended
as a polynomial around a fixed point:
\eqn\potent{V(T, T^*) = V(T_c, T_c^{*}) + 1/2 V''(T_c,
T_{c}^*) 
(T *_{2} T^* - T_c *_{2} T_c^*) + \cdots}
where $ V''(T_c, T_{c}^*) = m^2$ represents the mass
of the open string 
tachyon.
This mass depends on the values of $B$ and $F_i$
fields. 
Even without knowing a precise form for the
dependence on the $F_i$ fields, 
we can go directly to the particular case 
 $\int_{S^2} F_1 = 0, \int_{S^2} F_2 =  1$ 
which corresponds to a zero flux on the $D5$ brane and
a unit flux on the $\bar{D5}$
brane. As explained in \refs{\msa,\msb}, in this case the
open string between the 
 $D5-\bar{D5}$ pair contains in its spectrum a
massless scalar field instead
of a tachyon, which was obtained by considering a
projection on the open string Chan-Paton factors.
In terms of the equation \potent, this means that 
\eqn\mass{ V''(T_c, T_{c}) = 0}
In order to avoid any confusion, we still denote the
scalar field by $T$ even though is not a tachyon anymore but
a massless scalar field.

Because we are on the limit of large non-commutativity
due to the presence of
large $B$ and $F$ fields, the equations of motion
become:
\eqn\eqm{ [A_1^{\nu},[A_{1 \nu},A_{1 \mu}]] =
A_{1 \mu} *_{1} T *_{2} T^{*} - T *_{2} A_{2 \mu}
*_{2} T^* 
+ T *_{1} T^* *_{1} A_{1 \mu} - T *_{2} A_{2 \mu}
*_{2} T^*,}
\eqn\eqma{ [A_2^{ \nu},[A_{2 \nu},A_{2 \mu}]] =
A_{2 \mu} *_{2} T^* *_{1} T  - T^* *_{1} A_{1 \mu}
*_{1} T 
+ T^* *_{1} T *_{2} A_{2 \mu} - T^* *_{1} A_{1 \mu}
*_{1} T,}
\eqn\eqmb{-A_{1 \mu} *_{1} A^{ 1 \mu} *_{1} T + 2
A_{1}^{\mu} *_{1} T 
*_{2} 
A_{2 \mu} -
T  *_{2} A_{2}^{\mu} *_{2} A_{2 \mu}  = V'(T, T^*).}

In \refs\hkl, a solution has been given to these
equations for the case of vanishing $F_i$ fields. 
In our case we have
non-vanishing and non-equal $F_i$ fields 
so that solution is not valid for the new conditions described in the present 
paper. 

In order to obtain a solution, we will use a generalization of the results of
\refs\otw\ to the case of non-commutative theories. The system of 
 $D5-\bar{D5}$ branes is a triple 
 $(E_1, E_2, T)$ where $E_1, E_2$ are 
 $U(1)$ bundles over $S^2$ and the tachyon $T$ is a
map between them. 
In order to have a stable 3-brane
the condition on the bundle charges is:
\eqn\cc{c_1(E_2) - c_1(E_1) = 1}    
which is identical to the condition $\int_{S^2} F_1 = 0, \int_{S^2} F_2 = 1$ 
that we have in our paper. There is actually a minus sign difference between 
the equation \cc\ and the condition which appears in \otw. This is because 
the Chern-Simons terms contain terms like $B - F$ for the $D5$ brane and 
 $- (B - F)$ for the $\bar{D5}$ brane, our $c_1(E_i)$ will be their
 $- c_1(E_i)$. Therefore our condition \cc\ is the same as the condition of
stability of the $D5-\bar{D5}$ pair. The
equations of motion \eqm\ - \eqmb\ 
are  implied by the condition that all the fields be
holomorphic meaning that :
\eqn\hol 
{F_{i, z z} = F_{i, \bar{z} \bar{z}} = D_{\bar{z}} T =
0}

The holomorphicity of $T$ implies:
\eqn\holo{\bar{\partial}T + iA_{1,\bar{z}} *_{1} T -
i T *_{2} A_{2 \bar{z}} = 0 \ .}
In the large non-commutativity limit, the derivative
term can be 
neglected and 
the equation \holo\ becomes
\eqn\holoa{A_{1,\bar{z}} *_{1} T = T *_{2} A_{2 \bar{z}} \ .}
The fact that $E_1$ is trivial means that $A_1$ can be gauged to zero 
at infinity and because $E_2$ has $c_1(E_2) = 1$ means that $A_2$ can be
gauged to the form  $A_2|_\infty = \partial \theta$ where 
 $z = r e^{i \theta}$.
But equation 
\holoa\ is valid everywhere which means that $A_1, A_2$ have the same
behavior at infinity and this is contrary to the above choice for the
 assymptotic values at infinity. 
The only choice to satisfy equation \holoa\ 
is to consider a solution for $T$ as  $T|_\infty = 0$. 

In the case of zero $B$ field, the solution found in \refs\otw\ is of the form
 $T = f(r) e^{i \theta}$ with $f(0) = 0$ and $f(\infty)$ equal to a constant
which is actually the mass of the tachyon. This is the solution of the vortex
equations implied by the homophicity \hol\ and without neglecting the 
derivative terms. 
In our case, we do not change the condition $T|_0 = 0$ 
but the condition at $\infty$ is different, i.e. $T|_\infty = 0$. By changing 
the boundary condition at infinity we change the solution and we allow the
 system to have a trivial $T = 0$ solution everywhere. The change in the
boundary condition at infinity can be also seen from the fact that in 
\refs\otw\ the solution is $f(\infty) = \alpha$ where $\alpha$ is the mass of
the tachyon and this becomes zero for our specific choice for the fluxes.

Our solution
can be obtain by starting either with the solution of \refs\hkl\ and turning
on different fluxes on the $D5 - \bar{D5}$ pair or starting with the solution 
of \refs\otw\ and turning on a $B$ field. The first solution is a Gaussian one
which goes to zero at infinity and has a maximum at the origin. In one turns
now different values for the $F$ fields on the pair, the system is not
stable at the origin unless the tachyon solution becomes zero there so there 
is a factor $a(F_1, F_2)$ in front of the Gaussian solution which becomes zero
for  $\int_{S^2} F_1 = 0, \int_{S^2} F_2 = 1$. 
The second  solution has a zero at the origin and has a 
non-zero value at infinity. If one turns now a $B$ field, the value at zero is
unchanged but the value at $\infty$ is changed due to the previous
argument concerning the behavior and becomes zero for very large $B$ field. 
Therefore we see that it is either the
Gaussian solution which is deformed to a trivial one by turning on $F$ fields
or the vortex solution \ which is deformed to a trivial one by
turning on $B$ fields. We can navigate between the two non-trivial solutions
by switching on and off the $B$ field and the fluxes $F_i$. 

We have thus argued that in the case of large non-commutativity, 
with a trivial $E_1$ bundle and with $c_1(E_2) = 1$, there is a trivial
solution for the tachyon field $T = 0$. The discussion was based on the fact 
that we could neglect the derivative terms and the fact the tachyon 
field becomes massless for a specific choice for the values of the fluxes. 

\newsec{Fractional Branes and Conifolds}

We will use the result of the previous section to discuss the
conjecture stated in \refs\dmfrac. 
In the case of a conifold singularity, there is an apex where blowing up a
vanishing $S^2$ cycle resolves the singularity. If we probe the singularity
with $D3$ branes, we can study aspects of the 4 dimensional 
field theories on the worldvolumes of the $D3$ branes. There are two types of 
 $D3$ branes in the theory, the integer (anti) 
$D3$ branes which are orthogonal to
the singularity and fractional (anti) $D3$ branes which are (anti) $D5$ 
branes wrapped on the vanishing cycle \refs{\gubkleb,\dmfrac,\klenek,\kletse,
\ot}.  

The question is what happens when one wraps a $D5$ and an anti $D5$ on the
vanishing 2-cycle. In the usual case,
without fluxes, the $D5 - \bar{D5}$ system would just
annihilate by the usual 
tachyon condensation according to the formula:
\eqn\cond{
V(T_0) + 2 M_{D5} = 0}
where $T_0$ is the expectation value for the tachyon.
This is a general formula expected to be valid for $D5 - \bar{D5}$
pair on any background by using the universality argument. 
It is only valid for the case when the fluxes on the branes are equal
and it is not expected to be valid for different fluxes as explained in the 
introduction. 

In the present paper we are in the case of different magnetic fluxes.
In \refs\dmfrac , it has been stated that if besides the
 $B$ field one turns on a unit magnetic flux on the
 $\bar{D5}$ brane and no magnetic flux on the  $D5$ brane, the result is that
 one has stable BPS integer $D3$ brane instead of tachyon condensation. 
 Their result
 was based on the fact that the correct gauge invariant quantity on a brane is
 $B_{NS} - F$ and that is what appears in the Chern-Simons terms. 
In \refs{\msa,\msb} the result is that in the above specified condition, 
there is masless scalar in the spectrum of the open string connecting the
  $D5 - \bar{D5}$ pair and this is the tachyon 
whose mass has become zero due to
 the presence of the $B$ fields and $F_i$ fields. 
If we are to obtain an integer $D3$ brane, this field should not appear 
in the spectrum. But this exactly what we have discussed in the previous
 section, i.e. the fact that the equation of motion (or the homorphicity 
conditions) imply that it exists a trivial solution $T = 0$. 
Therefore our above results are in complete agreement with
the expectations for the field theory on a integer $D3$ brane at a conifold 
singularity.
The tension of the $D5-\bar{D5}$ system wrapped on the
vanishing $S^2$ cycle is recovered as the tension of the integer $D3$ brane. 

To discuss the charge of the stable system one uses the form of the 
Chern-Simons term worked out in \refs\kw\  and which can be extended to a 
non-commutative field theory in the lines of \refs\ms. 
There are two terms in the Chern-Simons coupling, one involves the tachyon
field and is given by
\eqn\first{ \int C \wedge d\, {\Tr}\left(T\wedge D \bar{T}\right)\ .
}
and this is zero for our tachyon field solution.
The second term arises from the coupling of the $D3$ brane RR potential to the
relative $U(1)$ gauge field strength on the $D5-\bar{D5}$ pair, i.e.
 $\int C_4 \wedge (F_2 - F_1)$.  This means that in our case the induced 
 $D3$ brane charge is one as expected. Our solution is different
from the one of \refs\hkl\ where the first term in the Chern-Simons couplings 
contributed and the second was zero because $F_2 - F_1 = 0$ in their case.

What happens with the $U(1) \times U(1)$ gauge group
which existed on
the worldvolume of the initial $D5-\bar{D5}$ pair?   
The DBI coupling
\eqn\dbi{
\int~*F \wedge B = \int~F^{ab}B_{ab}}
tells us that the tachyon is charged under the
relative gauge group 
$A_-= A_1-A_2$
and is neutral under $A_+=A_1+A_2$, where $A_i, i=1,2$
are the 
gauge fields on the
$D5$ and $\bar{D5}$ respectively. In the usual case
when the 
tachyon condenses $A_-$ becomes massive and therefore
decouples from 
the
low energy spectrum by Higgs mechanism. The tachyon field plays the
role of the Higgs field and the Higgs mechanism appears via tachyon 
condensation. 
The other gauge field $A_+$, under which the tachyon
is neutral, gets confined\refs{\senworld,\yi,\hori}. In
our case the
Higgs mechanism does not occur because the tachyon
scalar field has a zero expectation value and therefore the combination of the
gauge fields remains massless and is not removed from the low-energy spectrum.
By using standard electric-magnetic duality as in \refs\hori, one can show
that the second gauge group does not become confined so both gauge groups
survive and this implies that the entire $U(1)\times U(1)$
gauge group survives. This is just the gauge group on an integer $D3$
brane at a conifold singularity 
The theory on the $D3$ branes is a commutative one because there is no 
 $B$ or $F$ fields along its worldvolume directions.

The above discussion tells us that a  $D5-\bar{D5}$
pair on a vanishing $S^2$ cycle with $F_1 = 0, F_2 = 1$ 
magnetic fluxes on the branes is a stable system and
is an integer $D3$ brane. 

In the presence of other $D3$ branes at the conifold
singularity, 
the overall gauge group will be changed by a factor
$U(1) \times U(1)$ 
which agrees with previous known results
\refs{\klebwit,\dmfrac}. 

Our discussion can be easily extended for orbifolded 
conifolds to describe pairs of $D5-\bar{D5}$ branes wrapped on different
vanishing 2-cycles
described in \refs{\ot}. 
 
\bigskip\medskip

\noindent{\bf Acknowledgments:} 
\bigskip
We would like to thank K. Dasgupta and S. Mukhi for
collaboration 
and critical readings of the manuscript and
A. Jevicki for valuable discussions. We 
would like to thank the Institute for Advanced 
Studies for hospitality.

\listrefs    
\end